\documentstyle[emulateapj,apjfonts,epsf]{article}

\lefthead{COLPI, GEPPERT, PAGE \& POSSENTI}
\righthead{CHARTING THE TEMPERATURE OF THE HOT NEUTRON STAR IN A SXRT}
\def \ergs {\rm {erg\,s^{-1}} }
\def \Lq {L_{\rm {q}}}
\def \Lout {L_{\rm {out}}}
\def \Qnuc {Q_{\rm {nuc}}}
\def \Tc {T_{\rm {c}}}
\def \Tcore {T_{\rm {core}}}
\def \trec {t_{\rm {rec}}}
\def \tout {t_{\rm {out}}}
\def \gcm {\rm {g\,\, cm^{-3}} }
\def \msuny {M_{\odot}\,\,\rm{yr}^{-1}}
\def \mdot {\dot {M}}
\def \Mcr {M_{\rm cr}}

\def \mdotav {\langle \dot {M}\rangle }

\newcommand{\be}{\begin{equation}} 
\newcommand{\ee}{\end{equation}} 

\newcommand{\Msun}{\mbox{$M_{\odot}\;$}}

\begin{document}
\title{Charting the temperature  of the Hot Neutron Star in a Soft X-ray Transient}
\author{Monica Colpi$^1$, Ulrich Geppert$^2$, Dany Page$^3$, and Andrea Possenti$^4$}
\affil{$^1$Dipartimento di Fisica, Universit\'a degli Studi di Milano Bicocca,
Piazza della Scienza 3, 20126 Milano, Italy; colpi@uni.mi.astro.it \\
$^2$ Astrophysikalisches Institut Potsdam, An der Sternwarte 16, D-14482, 
Potsdam, Germany; urme@aip.de \\
$^3$ Instituto de Astronom\'{\i}a, UNAM, 04510 M\'{e}xico D.F., M\'{e}xico;
page@astroscu.unam.mx\\
$^4$ Osservatorio di Bologna, Via Ranzani 1, 40127 Bologna, Italy
} 

\begin{abstract}

We explore the thermal evolution of a neutron star undergoing episodes of intense
accretion,  
separated by long periods of quiescence.
By using an exact cooling code we follow in detail the flow of heat in the star due
to the time-dependent accretion-induced heating from pycno-nuclear reactions in the 
stellar crust, to the surface  photon emission, and the neutrino cooling.
These models allow us to study the neutron stars of the Soft X-Ray Transients.

In agreement with Brown, Bildsten and Rutledge (1998) we conclude that the soft component
of the quiescent luminosity of Aql X-1, 4U 1608-522, and of the recently
discovered SAX J1808.4, can be understood as thermal emission from a cooling neutron
star with negligible neutrino emission.
However, we show that in the case of Cen X-4, despite its long recurrence time, strong
neutrino emission from the neutron star inner core is necessary to understand 
the observed low ratio of quiescent to outburst luminosity.
This result implies that the neutron star in Cen X-4 is heavier than the one
in the other systems and the pairing critical temperature $\Tc$ in its center must be
low enough (well below $10^9$ K) to avoid a strong suppression of the neutrino emission.

\end{abstract}


\section{INTRODUCTION}

Neutron stars in Soft X-Ray Transients (SXRTs) undergo recurrent surges of activity
separated by long phases of relative "quiescence".  
The detection of type I bursts during the  X-ray brightening observed in 
outburst ($\sim 10^{37-38}\ergs$) unambiguously indicate that episodes of intense
accretion occur onto the stellar surface.
The origin of the faint X-ray emission, observed
in quiescence, at the level of $\sim 10^{32-33}\ergs,$ is instead uncertain
and may result  either from 
surface thermal emission (Brown et al. 1998\markcite{BBR98})
and/or from shock emission due to an hidden 
fastly spinning pulsar 
(Stella et al. 1994\markcite{Setal94}; Campana et al.
1998a,b\markcite{Cetal98a}\markcite{Cetal98b}).

Brown, Bildsten and Rutledge (1998\markcite{BBR98}, BBR98) showed, using simple but concrete arguments, that
the "rock-bottom" emission, in quiescence, could result from the cooling of the neutron star made hot during
the events of intense accretion: the inner crust compressed by the loaded material becomes the site of
pycno-nuclear reactions that may deposit enough heat into the core to establish a thermal luminosity 
\begin{equation}
\Lq\sim {Q_{\rm {nuc}}\over m_u} \mdotav \sim 6\,\times 10^{32} 
{\mdotav\over 10^{-11} \Msun \rm {yr}^{-1}}
\,\,\,\ergs 
\end{equation}
detectable when accretion halts ($\mdotav$ is the time averaged accretion rate,
$Q_{\rm {nuc}}\sim 1.5$ MeV, the nuclear energy deposited per baryon, and $m_u$ the atomic mass unit).
Following this consideration, Rutledge et al. (1999\markcite{Retal 99}, 
2000\markcite{Retal00}) showed that spectral fits with accurate hydrogen 
atmosphere models lead to emitting areas consistent with a neutron star surface, 
pointing in favor of a thermal origin of the quiescent emission (or of part of it). 

The interpretation of the quiescent emission seen 
in SXRTs as due to ``cooling'' gives
the opportunity of probing the physics of the neutron star 
interior in an unprecedented way.
We study in detail the thermal evolution of transiently accreting  
sources with a full cooling code, computing the ``exact'' quiescent luminosity 
and relating it to the accretion history.
We then explore the consequences of the cooling hypothesis to infer properties of the 
underlying neutron star.

\section{THE COOLING MODEL}

We use an ``exact'' cooling code which solves the equations of heat transport and
energy conservation in a wholly general relativistic scheme (Page 1989\markcite{P89}).
The cooling sources are neutrino emission in both the crust (electron-ion 
bremsstrahlung, plasma and Cooper pair formation) and the core (modified Urca and
neutron/proton bremsstrahlung, Cooper pair
formation, and direct Urca in some cases), and surface photon emission. 
The heating source is the accretion-induced production of nuclear energy. 
The energy  released in the upper atmosphere by the infalling matter and in the envelope 
by thermonuclear burning is here neglected since it is rapidly radiated away
(Fujimoto et al 1987\markcite{Fetal87}; Brown 2000\markcite{B00}). 
The heat sources located in the crust are instead important (BBR98\markcite{BBR98})
and describe the readjustment, 
to chemical equilibrium, of the matter processed during the heavy accretion events.
Energy release by electron captures and neutron emissions is included following
Haensel \& Zdunik (1990a\markcite{HZ90a}) and proceeds at a rate proportional to the
instantaneous accretion rate $\mdot(t)$.
Pycno-nuclear fusions, releasing the bulk of the energy, continue when the star is
returning to its cooling state, since their characteristic time-scale
($\sim$ months) is comparable to or longer than the duration of a typical outburst.
This time lag is accounted for with a suitable heat release function
(see Possenti et al. 2001). The inclusion of all the non-equilibrium energy sources 
(down to $\rho = 1.137\times 10^{13}\,\,\gcm$) assumes that the entire stellar crust 
has been replaced; this would take $\sim 50$ million years for a time averaged accretion 
rate of 10$^{-11}\msuny$ (Haensel \& Zdunik 1990a\markcite{HZ90a}). 
The temperature gradients in the interior determine how much of this heat is stored and how
much directly flows away to the surface. We calculate the thermal conductivity using 
the results of Baiko and Yakovlev (1995\markcite{BY95}) 
in the crystallized crustal layers, Itoh et al. (1983\markcite{IMII83}),
Mitake et al (1984\markcite{MII84}), and Yakovlev (1987\markcite{Y87}) in the 
liquid ones, and Flowers \& Itoh (1980\markcite{FI81}) in the core.
The calculations by Potekhin, Chabrier and Yakovlev (1997\markcite{PCY97})
of the structure of envelopes made of accreted material are used to
relate the interior temperature, at a density of $10^{10}$ g
cm$^{-3},$
to the surface thermal luminosity $L_{\rm q}$, (i.e. the quiescent luminosity
visible in between outbursts) and to the corresponding effective 
temperature $T_{\rm eff}$.

{
\plotone{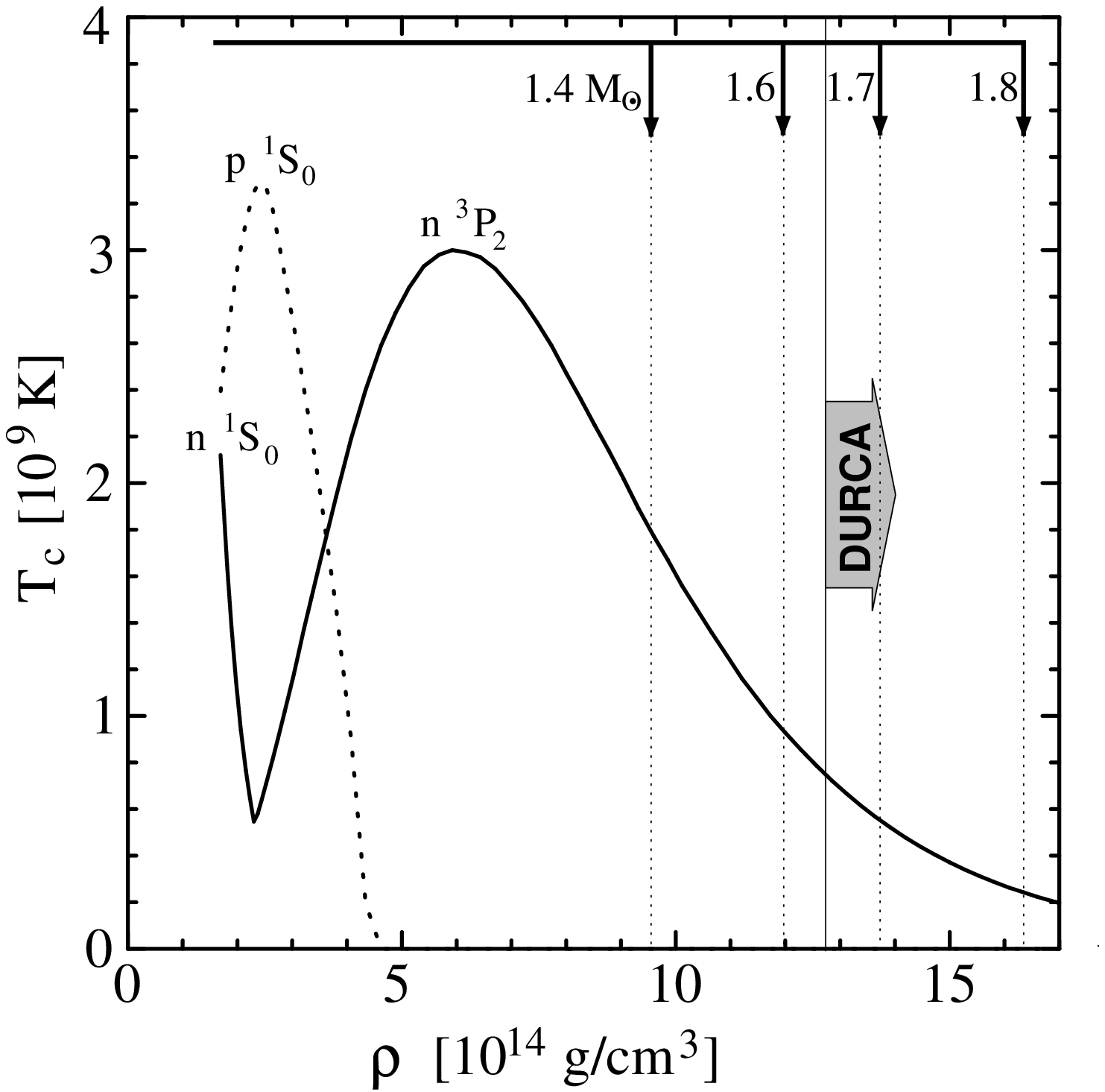}
\figcaption{\label{fig:pairing}\footnotesize
{Critical temperatures $\protect\Tc$ for both proton $^1$S$_0$ and 
neutron $^1$S$_0$ and $^3$P$_2$ pairing assumed in this work.
The central densities of the stars of various mass we study are indicated
as well as the threshold for the direct Urca (``DURCA'') process.}}
}

The chemical composition and the equation of state (EOS) of the crust are the ones of
an accreted crust (Haensel \& Zdunik 1990b\markcite{HZ90b}) 
and the state of this matter as well as its specific heat are followed
during the evolution of the star from the liquid to the solid
phase (Slattery et al 1980\markcite{SDD80}) and the Debye regime
(Shapiro \& Teukolsky 1983\markcite{ST83}). For the core
we follow Prakash, Ainsworth \& Lattimer (1988\markcite{PAL88}) using  a compression 
modulus $K_0 = 240$ MeV and a symmetry energy dependence $\propto \rho^{0.7}$.
With this particular EOS, the direct Urca process is allowed at densities
$\rho > 1.28 \times 10^{15}$ g cm$^{-3}$, which are attained in neutron stars 
heavier than a critical mass $\Mcr=1.65$ \Msun.  

Finally, we include the strong suppressing effects of superfluidity (neutron) and
superconductivity (protons) on both the specific heat and neutrino emission:
given the enormous uncertainty (Baldo et al 1998\markcite{BEEHJS98})
on the value of the critical temperature $\Tc$ for $^3$P$_2$ neutron pairing we plot 
explicitly in Fig.~\ref{fig:pairing} the values we adopt here.
We will refer to our 1.4 and 1.6 \Msun stars  as slow cooling neutron stars
and the 1.7 and 1.8 \Msun ones as fast cooling stars: 
the former ones have only the modified Urca process allowed in their core and suppression 
by neutron pairing is strong while the latter ones have the direct Urca process operating 
with suppression at lower temperatures so that fast neutrino cooling does affect 
their thermal evolution (Page 1998\markcite{P98}).


\section{TOWARD THE EQUILIBRIUM STATE}

As a preliminary step we verify whether transient accretion on an initially
cold neutron star is able to bring it to a hot stationary state and how fast
this happens.

{\medskip
\plotone{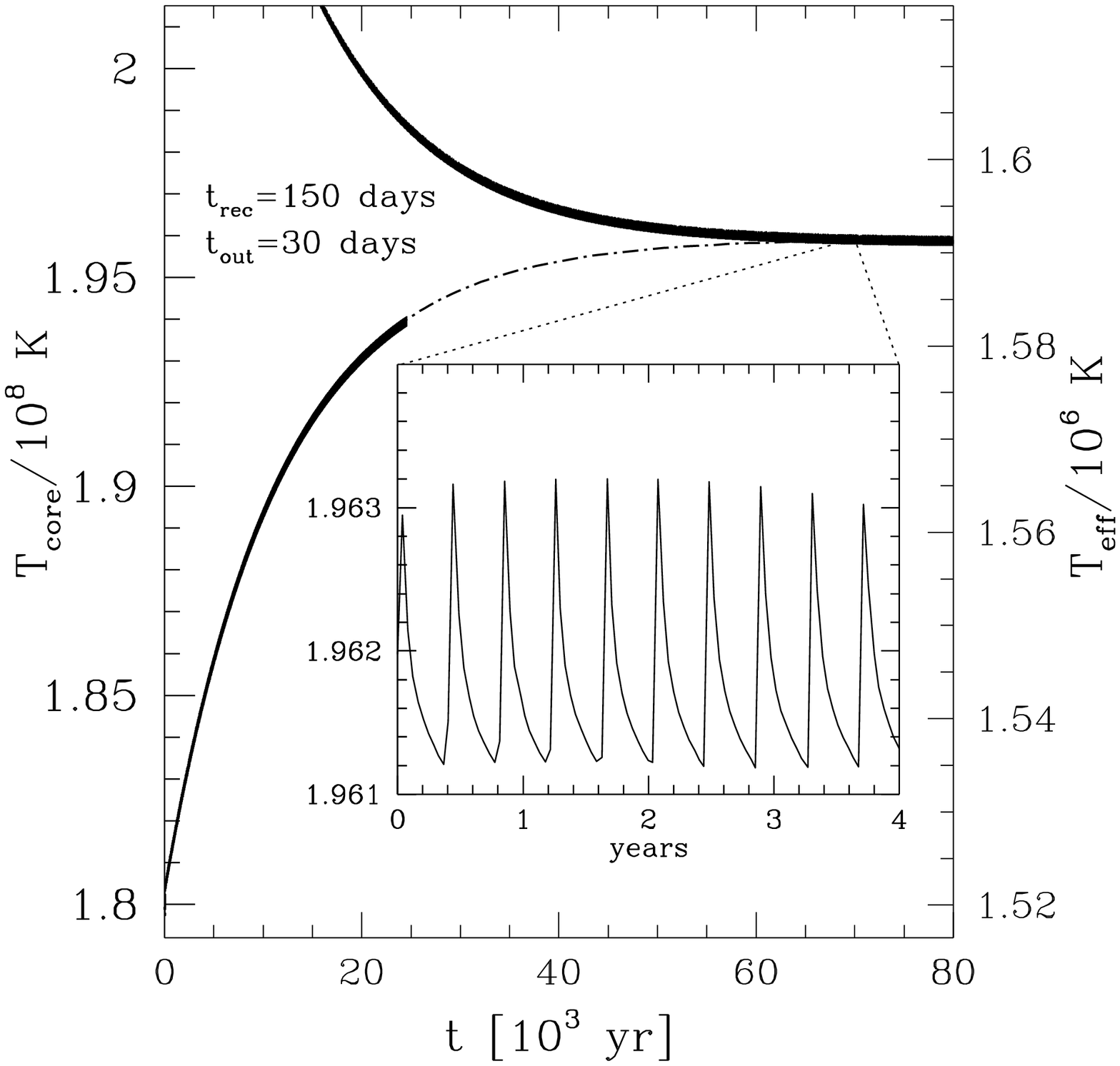}
\figcaption{\label{fig:asymptotic}\footnotesize
{Redshifted core temperature $T_{\rm core}$ ($10^8~K$) vs time 
$t$  soon after the onset of transient accretion, set at $t=0$. The right
scale reports the effective surface temperature as measured at infinity.
The solid lines refer to a 1.4 $\Msun$ superfluid star with transient 
accretion, $\Delta M=6\times 10^{-11}\Msun$, $\trec=150$ days, and $\tout=30$ 
days. The upper curve shows the evolution 
from a hotter state that
may result from an early phase of steady accretion, whereas the lower curve
mimics a resurrecting accretion period following a phase of pure cooling. 
The thickness of the lines
results from the rapid variations of the temperature due to accretion,
as can be seen in the insert where  $T_{\rm core}$ is plotted against 
time over a few cycles. 
The dot-dashed line shows the evolution of the same star 
with continuous accretion at 
$\langle\dot M\rangle=1.46\times~10^{-10}\,\,\Msun\,\,{\rm yr}^{-1}.$}}
\medskip}

For this, we model the accretion rate $\dot{M}(t)$ (as measured at infinity)  
with a fast exponential rise, on a time scale $t_{\rm rise}$, reaching a maximum 
$\dot{M}_{\rm max}$ followed by a power law decay of index $\alpha=3.$
Accretion is never turned off but becomes rapidly negligible in the decay phase. 
A new exponential rise recurs every $t_{\rm rec}$. 
$t_{\rm rise}$ is set by the duration of the outburst phase $t_{\rm {out}} 
(=3t_{\rm {rise}})$ which, for SXRTs is $\sim$30 days long.
The total accreted mass during one cycle,
$\Delta M=1.08 \; t_{\rm rise}\,\dot{M}_{\rm max}$,
is the key parameter and is used to determine $\dot{M}_{\rm max}$.
We define the outburst luminosity (at infinity) $L_{\rm out}$ such that the fluence is 
$L_{\rm out} \; t_{\rm out} =\eta \; \Delta M \,c^2$ 
with an efficiency 
$\eta=1-{\rm{e}}^{\phi}$ (${\rm e}^{\phi}=(1-2GM/Rc^2)^{1/2}\sim 0.15-0.30$
is the redshift factor). The time averaged accretion rate is 
$\langle\dot M\rangle=\Delta M/t_{\rm rec}.$

We show in Fig.~\ref{fig:asymptotic} an example of the heating (cooling) of an
initially cold (hot) neutron star after the onset of intermittent accretion. 
It is compared with a model having constant accretion at a rate equal to $\mdotav$. 
As expected, the star reaches thermal equilibrium and does it on a very short time 
scale $\tau_{\rm equ}\sim 10^{4}$ yr, much shorter than any binary evolution time, 
and, in particular, much shorter than the time necessary to replenish  the  crust
with fresh non-catalyzed matter. 
The equilibration  temperature is reached when the net injected  heat is
exactly balanced by the energy loss from the surface and/or from neutrino emission.  

\section{THE QUIESCENT LUMINOSITY}

Our stars reach a hot equilibrium state and glow, in quiescence, at a luminosity
$\Lq$ which varies with $\trec$ and $\Delta M,$ as illustrated in Fig.~\ref{fig:Lq-trec}. 
The solid lines refer to models with $\Delta M=6\times 10^{-11}\Msun$, 
compatible with that inferred for Aql X-1, 4U 1608-52, and Cen X-4. 
For a 1.4 \Msun star we have also drawn the curve corresponding to a lower value of
$\Delta M=10^{-11}\Msun,$ that may describe the fainter transient SAX J1808.4.
$\Lq$ depends crucially on whether fast neutrino emission
in the inner core is allowed or inhibited, as illustrated in Fig.~\ref{fig:Lq-trec}.

For slow cooling stars with superfluidity
(upper solid lines `1.4sf' and `1.6sf'), neutrino emission is totally suppressed
and the luminosity is determined by balance of nuclear heating with photon cooling. 
When superfluidity is not included, these stars are slightly less luminous 
as there is a small leakage of neutrinos due to the inefficient modified Urca process; 
the difference is very small and we omit the results in the figure. 

When fast neutrino emission is allowed (`1.7sf' and `1.8sf' lines),
the luminosity is much lower and the lowest $\Lq$ are obtained in the 
case neutrino emission is not affected by neutron pairing (`1.8nsf' case). 
In these fast cooling models the dependence of $\Lq$ on $\trec$ is much weaker than in the
slow cooling cases since most of the heat deposited during an accretion cycle is rapidly
lost into neutrinos. 
If $\Tc$ were very high at the 
center of even our most massive stars, then neutrino emission would be completely 
suppressed and the evolution of the superfluid 1.8 \Msun star would be almost identical 
to the superfluid 1.4 \Msun star. 
This is similar to what happens in cooling isolated neutron stars: 
the star's temperature is almost entirely controlled by the value of $\Tc$ 
(Page \& Applegate 1992\markcite{PA92}). 
We explicitly choose a relatively low $\Tc$ at high 
density to allow heavier stars (1.7 and 1.8 \Msun in our specific model) 
to have a qualitatively different behavior relative to the less massive ones. 

{
\medskip
\plotone{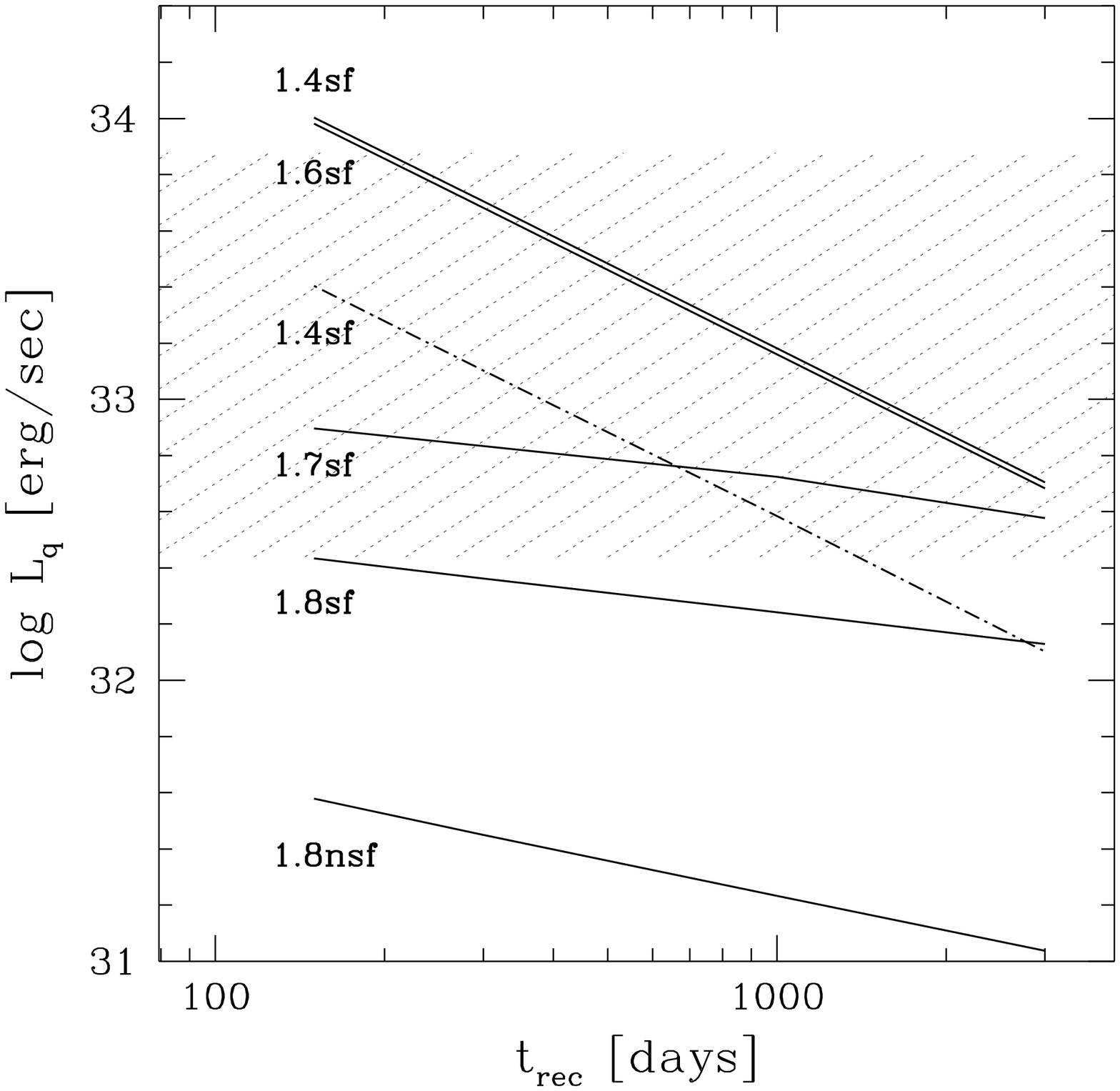}
\figcaption{\label{fig:Lq-trec}\footnotesize
{The (equilibrium) quiescent luminosity versus the recurrence time
resulting from our numerical calculations. The dashed
area covers the region of the observed luminosity in SXRTs. The solid lines
refer to a star accreting $\Delta M=6\times 10^{-11}\Msun$ in each
cycle. They are labelled by the mass of the star (in solar
masses) and the existence (sf) or nonexistence (nsf) of a superfluid phase.
The dot-dashed line shows our results for a 1.4 $\Msun$ superfluid
star loading $\Delta M=10^{-11}\Msun$ in each cycle.}}
\medskip}

\section{COMPARISON WITH DATA}

The most convenient quantity for comparison with data is the ratio $\cal R$ of 
fluence in quiescence $\Lq \,\trec$ to fluence in outburst 
$\Lout \,\tout \sim \eta \,\Delta M c^2$ (BBR98\markcite{BBR98}).
This gives us a direct measurement of the fraction $f$ of the heat released in the 
crust during accretion which, stored in the stellar interior, slowly 
leaks out to the surface 
\be
{\cal {R}} =  \frac{\Lq \,\trec}{\Lout \,\tout}
           =   f \;\;\; \frac{\Qnuc}{m_u \,c^2} \; \frac{{\rm e}^{-\phi}}{\eta} \;\; .
\label{equ:calF}
\ee
Our exact cooling code computes $\Lq$ and thus predicts $f$. 
This is illustrated in Fig.~\ref{fig:comparison} where we give model values for 
$\Lq/\Lout$ against $\trec/\tout$ and where we draw the theoretical line $f=1$ 
of Eq.~\ref{equ:calF} for 1.4\Msun and 1.8\Msun stars. 
Our numerical results when neutrino emission is negligible
(``1.4sf'' solid line) are very close to the $f=1$ line
showing a heat storage efficiency almost independent of $\trec/\tout$.
The ratio $\Lq/\Lout$ is dramatically reduced when direct Urca emission switches on, 
as in the core of our heavier stars (``1.7sf'' and ``1.8sf'' solid lines). 
In these cases, the heat storage efficiency decreases with decreasing
$\trec/\tout$ (along a line of given mass):
the neutron star in a transient with shorter $\trec$ has a warmer core and an enhanced 
neutrino emissivity ($\propto \Tcore^6$) leading to a lower value of $f$.

{
\medskip
\plotone{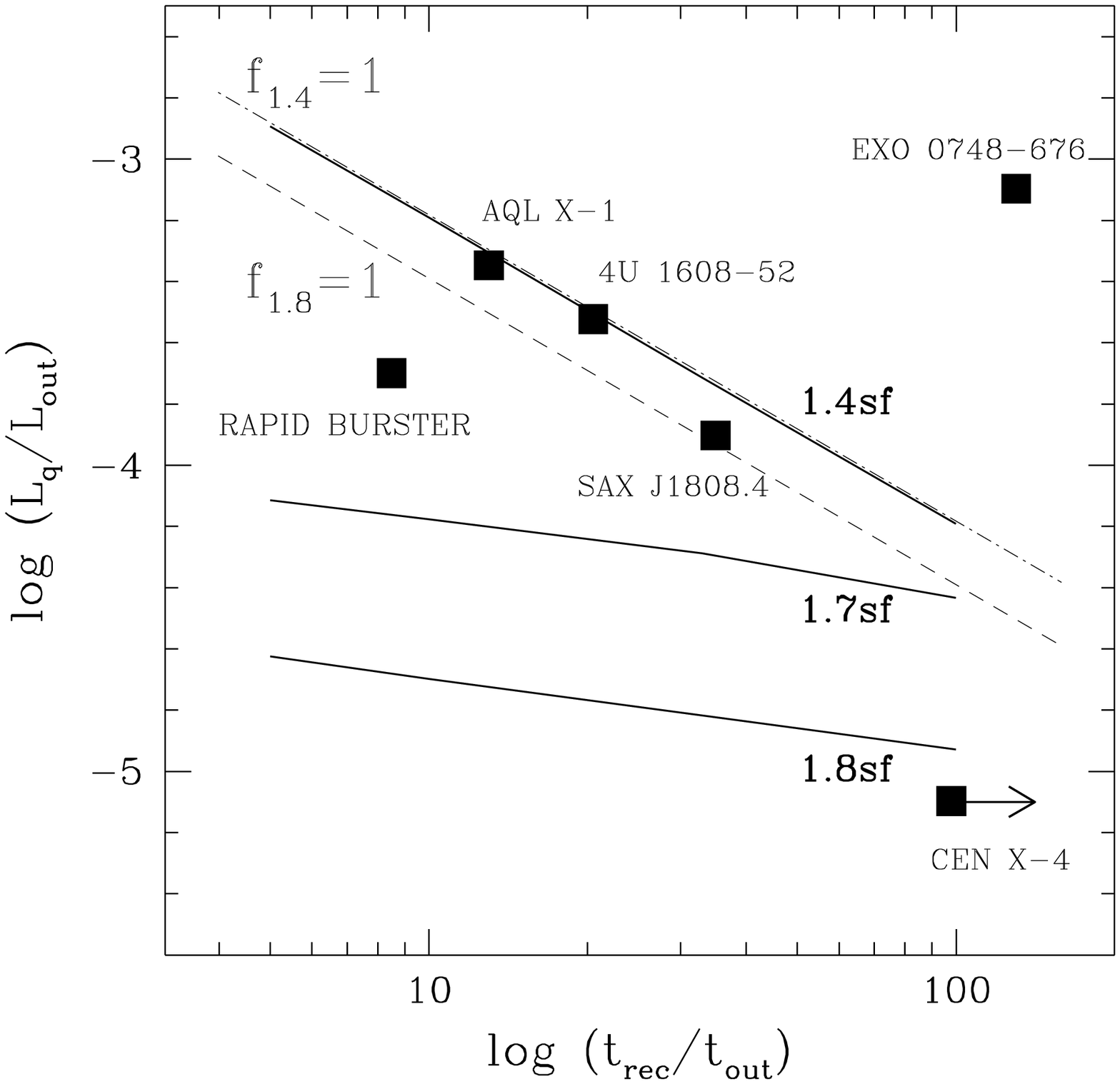}
\figcaption{\label{fig:comparison}\footnotesize
{Quiescent to outburst luminosity ratio plotted versus the ratio
of the recurrence time over the outburst time. 
Bold solid lines are the results of our numerical calculations as in
Fig.\ref{fig:Lq-trec}. 
The expected results for $f =1$, see Eq.~\protect\ref{equ:calF}, are shown for our 1.4 \Msun 
star (dashed-dot line) and 1.8 \Msun star (dashed line).
The filled squares represent the observed values.
}}
\medskip}

Rutledge et al. (2000) analyzed the observation of five SXRTs, 
Aql~X-1, Cen~X-4, 4U~1608-52, EXO~0748-767, and the Rapid Burster, 
made during the periods of lowest emission. 
Fitting the spectra with  the H atmospheric models 
of Zavlin et al (1996\markcite{ZPS96})
they calculated the thermal luminosity 
$\Lq$ and the outburst time averaged luminosity $\Lout$. 
All five systems 
display similar peak luminosities and outburst times. 
For the fainter source SAX~J1808.4,  
we take $\Lq=2.5 \times 10^{32}\ergs,$ $\Lout=10^{36}\ergs,$ $\tout=20$ days and 
$\trec=700$ days (Stella et al. 2000\markcite{Setal00}).
Overlying all these observations in the diagram of Fig.~\ref{fig:comparison},
it appears remarkable that Aql~X-1 and 4U~1608-52 stands just on the theoretical line
constructed for 1.4 \Msun (close also to 1.6 \Msun as shown in Fig.~\ref{fig:Lq-trec}). 
The Rapid Burster would fit the theory if its mass lies in an intermediate range with
fast neutrino emission strongly controlled by pairing.
Cen~X-4 is adjacent to the 1.8 \Msun model star results, having an $f_{1.8}$ of 0.18.
Such a low efficiency means that about 80\% of the deposited heat is not observed
to flow back to the stellar surface:
only a very efficient {\em UnRecorded Cooling Agent} such as the direct Urca process
is able to provide such a strong energy drain.
Note however that the ratio $\trec/\tout$, for Cen~X-4, 
could be underestimated  
as only two outbursts have been recorded, in 1969 and 1979.
Thus, we can not exclude that Cen~X-4 is close  to 
the curve $f_{1.4}=1;$ this however requires a very large $\trec/\tout$ of 1,000.
Its low value of $\Lq/\Lout$ could alternatively indicate the presence of a not fully 
replaced crust.
We here take the view that Cen~X-4 is heavy.

Having calculated the upper theoretical bound on the efficiency of the 
rediffusion of the pycno-nuclear energy, we can interpret the peculiar behavior of 
EXO~0748-767 as due to an extra luminosity resulting from the gravitational energy 
released by a faint accretion and/or by the interaction of the infalling matter with the 
magnetosphere (BBR98\markcite{BBR98}).

Our results confirm the overall picture suggested by BBR98 in their parametric approach: 
the quiescent luminosity seen in SXRTs comes from the rediffusion toward the surface 
of the heat deposited in the core by the pycno-nuclear reactions triggered in the crust 
due to transient accretion.

\section{WEIGHTING NEUTRON STARS AND PROBING SUPERFLUIDITY} 

We can now use the established link of $f$ with mass, to infer properties of the star in a transient
through the comparison with the data.
The mass signature in the $\Lq/\Lout$ vs $\trec/\tout$ plot is mainly determined by the
state of the matter in the inner core of the neutron star.
The ratio $\Lq/\Lout$ has a clear drop when the source mass is above the critical mass   
for fast neutrino emission $\Mcr$ determined by the
value of the pairing critical temperature $\Tc$ in the innermost core.
The value  of $M_{\rm cr} = 1.65 \; \Msun$ used here is model dependent,
but combining future independent measurements of the mass 
with the observed ratio $\Lq/\Lout$ will open the possibility of
actually measuring the value of $M_{\rm cr}$ and/or also
infer an upper limit on the critical pairing temperature $\Tc$.
Notice, however, that many other fast neutrino emission channels, beside 
the simplest nucleon direct Urca considered here, are possible and
the constraints we could obtain about the value of $\Tc$ may then well refer to hyperon 
or quark pairing (Page et al. 2000\markcite{PPLS00}).

Awaiting future independent mass measurements, we can
discriminate between stars with masses close to the canonical value 1.35$-$1.40 \Msun
(Thorsett \& Chakrabarty 1999\markcite{TB99}) 
and more massive objects. 
In this context, the ratio $\Lq/\Lout$ of Cen X-4 can be explained
if its neutron star is heavier, and even probably much heavier, than in the other 
observed SXRTs. In low mass binaries, the neutron star could have undergone 
a substantial mass load (0.1 to 0.5~\Msun) while being spun up to very short periods.
Observational evidence (Casares et al. 1998\markcite{Cetal98}) and theoretical arguments
(Stella \& Vietri 1999\markcite{SV99}; Possenti et al. 1999\markcite{Petal99}) 
can support this view.

Charting the temperature of the old hot neutron star in a soft X-ray transient is hence
a valuable tool to investigate dense matter in neutron star cores which is complementary 
to the study of isolated young cooling neutron star.
It moreover opens the possibility to study the interior of stars which can be more
massive than the isolated ones.

\medskip

We thank L. Bildsten, G. Pavlov and R. Rutledge for a critical 
reading of the manuscript and the Institute of Theoretical 
Physics at Santa Barbara for its hospitality during the 
last phase of this work. This work was supported by ASI contract at Milano, 
Conacyt (grant \#27987E) and UNAM-DGAPA (grant \#IN119998),
a binational grant DFG (\#444 - MEX - 1131410) Conacyt at AIP and UNAM,
and the NSF under Grant \# PHY94-07194 at the ITP (UCSB).



\begin{references}

\reference{BY95}
Baiko, D. A., \& Yakovlev, D. G. 1995,
Astron. Lett., 21, 702

\reference{BEEHJS98}
Baldo, M., Elgaroy, O., Engvik, L., Hjorth-Jensen, M. \& and Schulze, H.-J. 1998,
Phys. Rev., C58, 1921


\reference{B00}
Brown, E. F. 2000,
ApJ, 531, 988

\reference{BBR98}
Brown, E. F., Bildsten, L., \& Rutledge, R. E. 1998,
ApJ, 504, L95 (BBR98)

\reference{Cetal98a}
Campana, S., Colpi, M., Mereghetti, S., Stella, L.,  \& Tavani, M. 1998a, 
A\&A Rev., 8, 279

\reference{Cetal98b}
Campana, S. et al. 1998b, ApJ, 499, L65 

\reference{Cetal98}
Casares, J., Charles, P., \& Kuulkers, E. 1998,
ApJ, 493, L39

\reference{FI81}
Flowers, \& Itoh, N. 1981,
ApJ, 250, 750

\reference{Fetal87}
Fujimoto, M. Y., Hanawa, T., Iben, I., \& Richardson, M. B. 1987,
ApJ, 315, 198

\reference{HZ90a}
Haensel, P., \& Zdunik, J. L. 1990a,
A\&A, 227, 431

\reference{HZ90b}
Haensel, P., \& Zdunik, J. L. 1990b,
A\&A, 229, 117

\reference{IKMS84}
Itoh, N., Kohyama, Y., Matsumoto, M., \& Seki, M. 1984,
ApJ, 285, 758

\reference{IMII83}
Itoh, N., Mitake, S., Iyetomi, H., \& Ichimaru, S. 1983,
ApJ, 273, 774

\reference{MII84}
Mitake, S., Ichimaru, S., \& Itoh, N. 1984,
ApJ, 277, 375

\reference{P89}
Page, D. 1989,
Ph.D. Dissertation, SUNY at Stony Brook

\reference{P98}
Page, D. 1998,
in The Many Faces of Neutron Stars,
eds. A. Alpar, R. Buccheri, \& J. van Paradijs
(Dordrecht: Kluwer Academic Publishers), 539

\reference{PA92}
Page, D, \& Applegate, J.~H. 1992,
ApJ, 394, L17

\reference{PPLS00}
Page, D., Prakash, M., Lattimer, J. M., \& Steiner, A. 2000,
Phys. Rev. Lett., 85, 2048

\reference{Petal99}
Possenti, A., Colpi, M., Geppert, U., Burderi, L., \& D'Amico, N. 1999,
ApJS, 125, 463

\reference{Pos01}
Possenti, A., Colpi, M., Page, D., \& Geppert, U. 2001, in Evolution of
Binary and Multiple Star Systems, to appear in ASP Conf. Ser.

\reference{PCY97}
Potekhin, A. Y., Chabrier, G., \& Yakovlev, D. G. 1997,
A\&A, 323, 415

\reference{PAL88}
Prakash, M., Ainsworth, T. L., \& Lattimer, J. M. 1988,
Phys. Rev. Lett., 61, 2518

\reference{Retal99}
Rutledge, R. E., Bildsten, L., Brown, E. F., Pavlov, G. G., \& Zavlin, V. E. 1999,
ApJ, 514, 945

\reference{Retal00}
Rutledge, R. E., Bildsten, L., Brown, E. F., Pavlov, G. G., \& Zavlin, V. E. 2000,
ApJ, 529, 985

\reference{ST83}
Shapiro, S. L., \& Teukolsky, S. A. 1983,
Black Holes, White Dwarfs, and Neutron Stars
(New York: Wiley-Interscience)

\reference{SDD80}
Slattery, W. L., Doolen, G. D., \& DeWitt, H. E. 1980,
Phys. Rev. A21, 2087

\reference{Setal94}
Stella, L., Campana, S., Colpi, M., Mereghetti, S.,  \& Tavani, M. 1994, 
ApJ, 423, L47

\reference{Setal00}
Stella, L., Campana, S. Mereghetti, S., Ricci, D., \& Israel, G. L. 2000,
ApJ, 537, L115

\reference{SV99}
Stella, L., \& Vietri, M. 1999,
Phys. Rev. Lett., 82, 107

\reference{TB99}
Thorsett, S. E.,\& Chakrabarty, D. 1999,
ApJ, 512, 288


\reference{Y87}
Yakovlev, D. G. 1987,
Sov. Astron., 31, 346

\reference{ZPS96}
Zavlin, V. E., Pavlov, G. G., \& Shibanov, Yu. A. 1996,
A\&A, 315, 141




\end{references}
\end{document}